# Graphene produced by radiation-induced reduction of graphene oxide

**Prashant Kumar, K. S. Subrahmanyam, C. N. R. Rao**[*]

Chemistry and Physics of Materials Unit, International Centre for Materials Science, New Chemistry Unit and CSIR Centre of Excellence in Chemistry, Jawaharlal Nehru Centre for Advanced Scientific Research, Jakkur P.O., Bangalore-560 064, India

**Abstract:**

Effect of irradiation on graphene oxide by sunlight, UV light and KrF excimer laser has been investigated in detail. Both sunlight and ultraviolet light reduce graphene oxide well after prolonged irradiation, but laser irradiation produces graphene with negligible oxygen functionalities within a short time. Laser irradiation is also useful for one-step synthesis of metal particle decorated graphene. Laser irradiation of graphene oxide appears to be an efficient procedure for large-scale synthesis of graphene.

[*]**Corresponding author:** Tel. +91 80 23653075; fax: +91 80 22082760.
E-mail address: cnrrao@jncasr.ac.in.

1. Introduction

Graphene has been a topic of intense research in the last four to five years.[1-5] Graphene was first prepared by mechanical exfoliation of graphite crystals[6] and this technique has been commonly employed in many of the studies. This method is not suitable for large scale production of graphene and there have been attempts to achieve this by employing various strategies such as by epitaxial growth on Si-terminated 6H-SiC (0001)[7], chemical vapour deposition,[8-10] electrostatic force assisted exfoliation[11] etc. The most sought after method is by the solution route[12-13], but there are problems faced with complete exfoliation of graphite by chemical means and the stabilization of exfoliated graphene in liquid suspension. A common procedure has been to reduce exfoliated graphene oxide by hydrazine hydrate[14]. Graphene samples obtained by such chemical methods generally have surface functional groups containiing oxygen. Electrochemical reduction[15] and localized reduction[16] by means of heated AFM tips have been attempted to overcome this problem. Electromagnetic radiation can, in principle, be used for reducing graphene oxide. Laser reductions of graphite oxide[17] and photothermal deoxygenation of graphene oxide[18] by camera flash have been reported recently.

We have attempted to explore the use of different irradiation sources to achieve the reduction of graphene oxide. For this purpose, we have employed[19] sunlight, ultraviolet radiation and a KrF excimer laser. We have found that irradiation of graphene oxide with the KrF excimer laser yields graphene nearly devoid of surface oxygen. Excimer laser irradiation of a solution containing graphene oxide and a metal compound such as $H_2PtCl_6$, yields graphene sheets decorated with metal particles in one step.

## 2. Experimental

Graphite oxide was prepared by the modified Hummer's method.[20] Graphite oxide readily forms a stable colloidal suspension in water. The aqueous suspension was subjected to ultrasonic treatment (300 W, 35 kHZ) to produce single-layer graphene oxide (GO). The GO solution taken in a petri dish was exposed to sunlight for a few hours. For treatment with ultraviolet radiation, the solution was irradiated with a Philips low-pressure mercury lamp (254 nm, 25 W, 90 µW/cm$^2$). A Lambda Physik KrF excimer laser (248 nm, 5 Hz) was employed to irradiate aqueous solutions of GO taken in a quartz vail. The aluminum metal slit (beam shaper) which usually gives a rectangular beam was removed during laser irradiation of the solution. This makes laser energy almost uniform throughout the area where graphene oxide is present. 300 mJ beam energy at 5 Hz reprate was used for the purpose. The reduced graphene samples obtained by irradiation with sunlight, ultraviolet light and KrF excimer laser are designated as SRGO, URGO and LRGO respectively. Apart from graphene oxide reduction by various irradiation sources, graphite oxide solution in water was directly irradiated with excimer laser for an hour to see whether sonication step can be avoided.

An aqueous solution of $10^{-4}$ M $H_2PtCl_6$ (0.083 ml) was mixed to a 1 ml of aqueous solution of GO (1mg/ml) and subjected to sonication followed by excimer laser irradiation. Excimer laser irradiation was carried out in a quartz vail at laser energy 300 mJ and for 18000 laser shots. Similarly, 0.08 ml of aqueous $10^{-4}$ M $H_2PdCl_4$ was mixed with 1 ml of GO (1mg/ml) for laser irradiation at the same energy (300 mJ) with 18000 laser shots.

Fourier transform infrared spectra were recorded using an IFS66v/s Bruker spectrometer. Samples for infrared measurements were prepared by first removing the water out of the solution. Field emission scanning electron microscopy (FESEM with a NOVA NANOSEM 600) and atomic force microscopy in dynamic force mode (Innova scanning probe microscope (SPM)) were carried out to look into the morphology of various samples. For FESEM, samples were drop-casted onto a silicon substrate and imaged in high vacuum mode. Raman spectroscopy was also carried out. Core-level X-ray photoelectron spectroscopy measurements were carried out (with Al $K_\alpha$ radiation, 1486.6 eV) using a VSW EA45 analyzer at 100 eV pass energy for dried solid samples of GO and LRGO. C 1s and O 1s peaks were analyzed. Photoluminescence (PL) spectra were recorded with a Perkin-Elmer model LS55 luminescence spectrometer with 325 nm excitation. Two probe resistance measurements were carried out on solid samples of GO, SRGO and LRGO using a digital multimeter.

3. **Results and discussion**

In Fig 1(a), we show a photograph of the GO solution in water medium which is brownish yellow in colour. It gradually turns reddish after 2 h of irradiation by sunlight as can be seen from Fig 1(b). It finally turns black in colour after 10 h of irradiation by sunlight as shown in Fig 1(c). The colour change of the graphene oxide from brownish yellow to black is clear evidence of the occurrence of reduction. From the IR spectra, we see that the intensity of the carbonyl stretching band decreases substantially after 10 hour irradiation (Fig 2). Similarly, the intensities of bands due to other oxygen containing functional groups also decrease after the prolonged sunlight irradiation. Figures 3(a) and (b) show FESEM images ofgraphene oxide before and after irradiation by sunlight (10 h). We notice only a marginal change in the surface after irradiation with only minor buckling of the flakes.

In Fig 4, we show the IR spectrum to demonstrate the effect of ultraviolet irradiation on the graphene oxide solution. We observe that the intensity of the carbonyl stretching band in the IR spectrum decreases markedly after ultraviolet irradiation for 2 h. Bands due to other oxygen containing functional groups disappear upon UV irradiation. Some changes are observed in the FESEM image of graphene oxide subjected to ultraviolet irradiation for 2 h (Fig 5). UV treatment seems to render the surface buckled. More graphene edges are visible on the surface after irradiation.

We have obtained the most interesting results with KrF excimer laser treatment. The brownish yellow colour of the graphene oxide solution becomes deeper brownish red even after short laser irradiation (approximately 4 minutes of laser treatment, 1000 laser shots at 300 mJ). After 1 h of laser treatment (18000 laser shots), the graphene oxide solution turns black in colour (Fig 6). The fast colour change obtained with the excimer laser is due to the large laser fluence as compared to the areal energy density for sunlight and ultraviolet light. We see in the IR spectra the near disappearance of the carbonyl stretching band as well as of other bands due to oxygen functionalities after 1 h of irradiation (Fig 7). Excimer laser irradiation not only reduces the graphene oxide effectively but also fragments the graphene as can be seen from Fig 8(a)-(c). Excimer laser irradiation of a solid graphene oxide disc gives rise to highly porous surface features with large number of edges as can be seen from the FESEM image in Fig 8(d). Solid graphene oxide requires only 4-5 laser shots (1 second time for 5 Hz reprate) to be completely reduced to black graphene, devoid of any oxygen containing functional groups. Such reduction of graphene oxide in a few seconds makes excimer laser an excellent choice for reduction. This is noteworthy since no reducing agent is required in this procedure and the extent of reduction is better than that by any other means. The wattage/cm$^2$ for KrF excimer laser and the low pressure

UV (254 nm) mercury lamp used in our experiment are 3 mW /cm$^2$ and 90 µW/cm$^2$ respectively. The areal power density of KrF excimer laser is thus approximately 33 times larger than that for the mercury lamp. X-ray photoelectron spectroscopy of solid graphene oxide and excimer laser reduced graphene oxide (LRGO) was also carried out and the signals of C 1s and O 1s were analyzed. We find the signals due to C=O and C-O at 286.5 eV and 285 eV are not observed in LRGO.

Raman spectra of the reduced samples such as LRGO show D, G and 2D bands respectively at 1332, 1580 and 2640 cm$^{-1}$. Atomic force microscopic image of the laser reduced samples shows a large number of tiny flakes (20-120 nm) as shown in Fig 9. Some large flakes also remain unbroken. The LRGO flakes are 1-7 layers in thickness as shown by AFM height distribution histogram, the average height being 1.2 nm.

GO shows a luminescence band in 550-600 nm region for 325 nm excitation. After irradiation by various irradiation sources viz. sunlight, ultraviolet light and excimer laser, the solution emits in the blue region, with a band centred around 430-440 nm as shown in Fig 10(a). The intensity of emission increases as LRGO > URGO > SRGO. The CIE chromaticity diagram of GO and reduced graphene oxide samples achieved by various treatments are shown in Fig 10 (b). A possible origin of the blue photoluminescence in reduced graphene oxide is the radiative recombination of electron–hole (e-h) pairs generated within localized states.

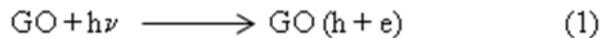

$$GO + h\nu \longrightarrow GO\,(h + e) \qquad (1)$$

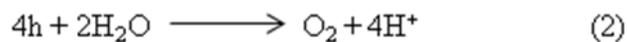

$$4h + 2H_2O \longrightarrow O_2 + 4H^+ \qquad (2)$$

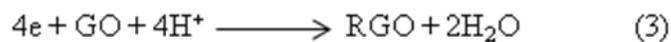

$$4e + GO + 4H^+ \longrightarrow RGO + 2H_2O \qquad (3)$$

The energy gap between the π and π* states generally depends on the size of $sp^2$ clusters or conjugation length. [21, 22] Interaction between nanometer-sized $sp^2$ clusters and finite-sized molecular $sp^2$ domains could play a role in optimizing the blue emission in irradiation reduced graphene oxide materials.

Electronic properties change after irradiation of GO as expected. Two-probe resistance measurement show that the resistivity of GO, SRGO and LRGO are 5000, 550 and 41 ohm-cm respectively. This is expected as reduced GO will have less oxygen containing functionalities which would make the samples more conducting. It may be noticed that Jung et. al.[23] have reported 4 orders of magnitude difference in the 4-probe electrical conductivity after the chemical reduction.

We have also carried out experiments on the irradiation of graphite oxide solutions in water by the excimer laser at 300 mJ beam energy. It is found that after 1 h of irradiation, the brownish yellow colour of graphite oxide solution changes to black as can be seen from Fig 11. Thus, the sonication step is actually not necessary because excimer laser itself can do fragmentation as well as reduction. It also demonstrates the generality of this technique.

When a solution of graphene oxide and $H_2PtCl_6$ in water was irradiated with excimer laser radiation at 300 mJ beam energy for 1 h at 5 Hz reprate, platinum nanoparticles formed and decorated the graphene sheets. We show the Pt nanoparticle decorated graphene in the FESEM image in Fig 12 (a), the platinum nanoparticles have a diameter in the 3-10 nm range. The energy-dispersive X-ray spectrum of platinum-decorated LRGO, given in the inset to Fig 12(a), clearly shows platinum peaks. Palladium nanoparticles of diameter 4-15 nm got deposited on graphene sheets by laser irradiation of a solution containing GO and $H_2PdCl_4$ as can be seen from the FESEM image in Fig 12(b). The energy-dispersive X-ray spectrum shown in the inset to Fig 12(b) reveals peaks due to palladium. This is a significant result in the sense that the laser can be employed for a one-step synthesis of metal nanoparticle decorated graphene sheets to accomplish the simultaneous reduction of graphene oxide and the metal salt.

4. **Conclusions**

Irradiation with different light sources has been employed to reduce oxygen functionalities of graphene oxide to produce graphene. In particular, the excimer laser radiation is potentially effective in preparing graphene nearly devoid of oxygen functionalities. Excimer laser can also be employed to achieve simultaneous reduction of graphene oxide and metal salts to produce metal particle decorated graphene. It is noteworthy that the reduced graphene samples emit blue light.

**Figure Captions**

**Fig 1.** Photographs of GO solutions (a) before irradiation by sunlight and after irradiation by sunlight for (b) 2 h and (c) 10 h.

**Fig 2.** IR spectra of (a) untreated GO and (b) after 10 h irradiation of sunlight.

**Fig 3.** FESEM images of (a) untreated GO and (b) after 10 h of irradiation by sunlight.

**Fig 4.** IR Spectra of (a) untreated GO and (b) after 2 h of UV irradiation.

**Fig 5.** FESEM images of (a) untreated GO and (b) after 2 h of UV irradiation.

**Fig 6.** Photographs of (a) untreated GO and after irradiation with an excimer laser radiation (300 mJ) for (b) 1000 shots and (c) 18000 shots.

**Fig 7.** IR spectra of (a) untreated GO and after laser treatment in liquid form with excimer laser at 300 mJ laser energy for (b) 1000 shots and (c) 18000 shots.

**Fig 8.** FESEM images of (a) untreated GO and after treatment of solutions with excimer laser radiation at 300 mJ laser energy with (b) 1000 shots and (c) 18000 shots and (d) the solid with 5 laser shots.

**Fig 9.** AFM image of laser reduced sample along with height distribution histograms.

**Fig 10.** (a) PL spectra of GO (1), SRGO (2) URGO (3), LRGO (4) and (b) CIE chromaticity diagram.

**Fig 11.** Photographs of (a) graphite oxide solution in water and of (b) excimer laser reduced graphite oxide.

**Fig 12.** FESEM images of metal nanoparticles decorated graphene sheets for (a) Pt and (b) Pd. Insets show energy dispersive x-ray spectra.

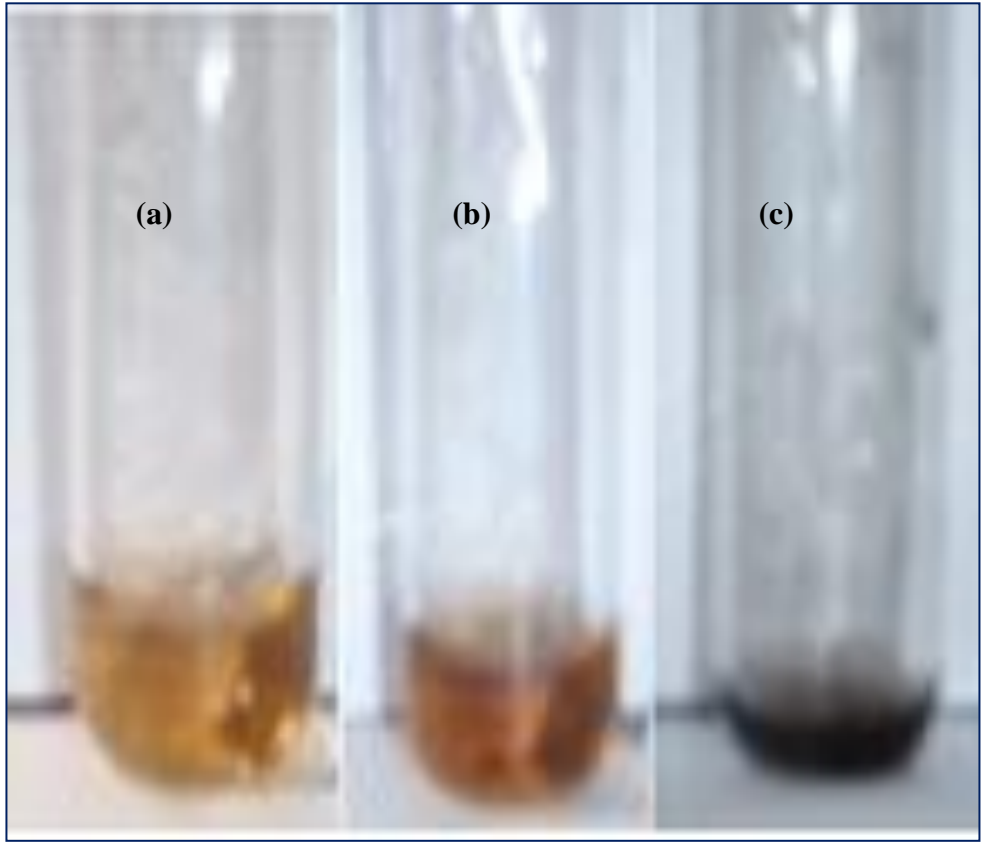

**Fig 1**

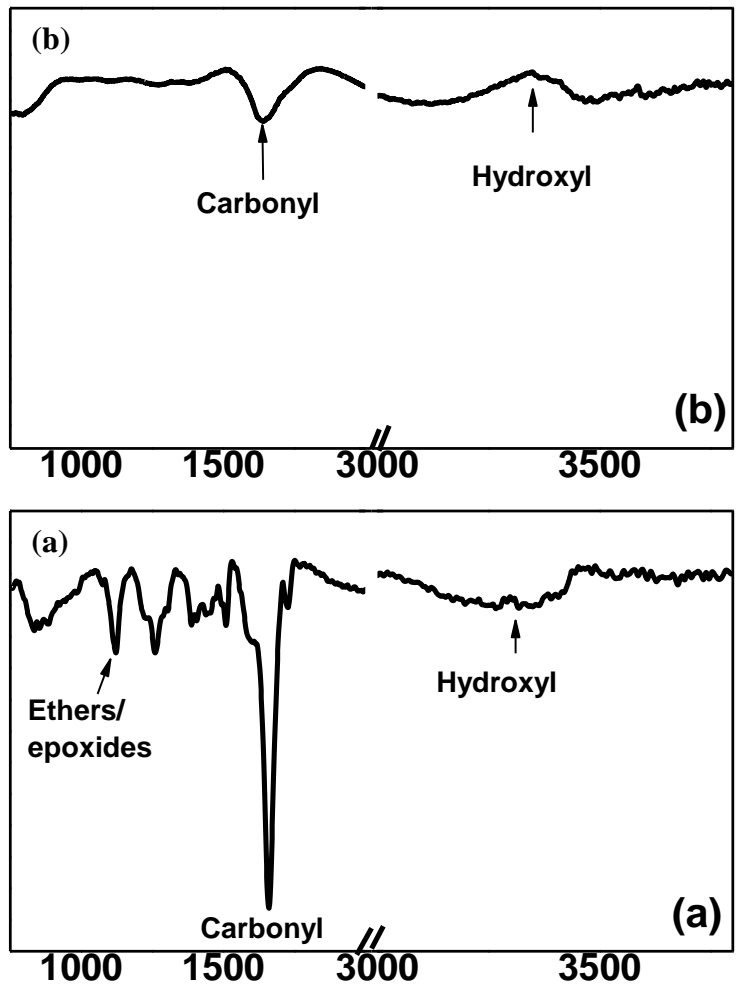

Fig 2

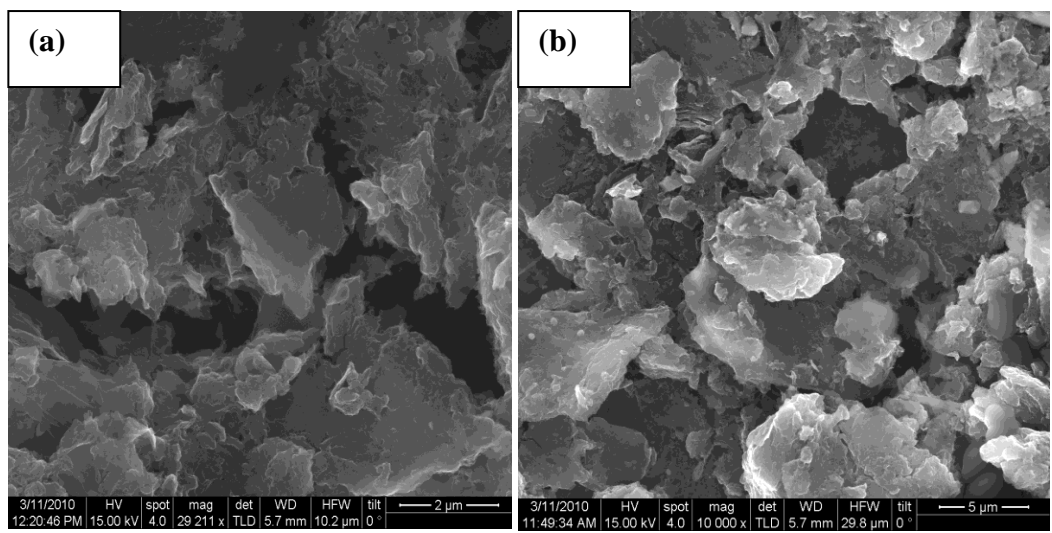

**Fig 3**

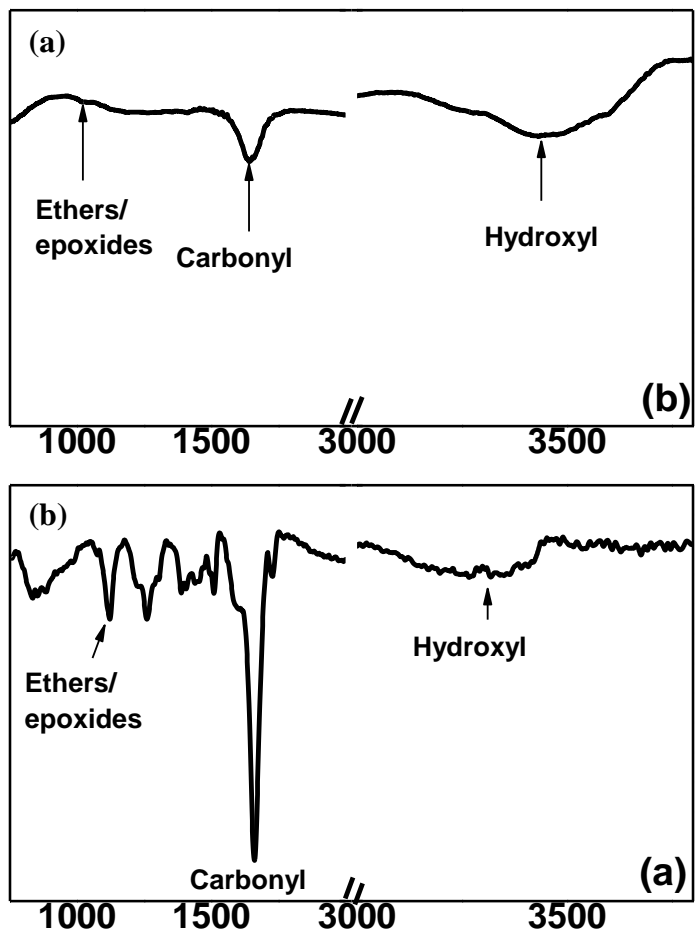

Fig 4

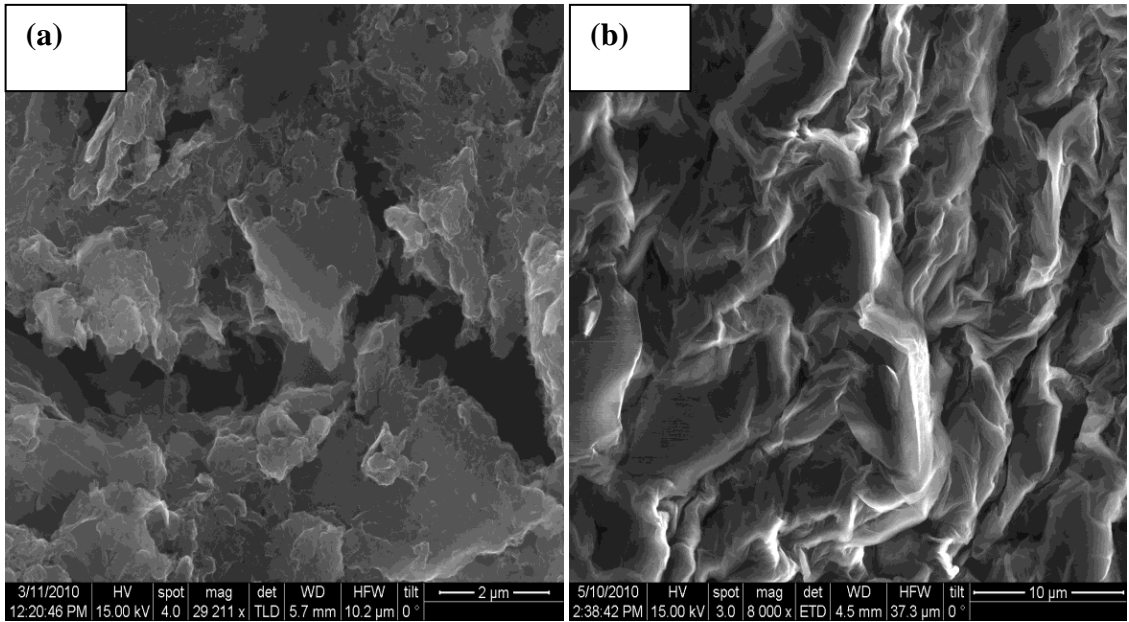

**Fig 5**

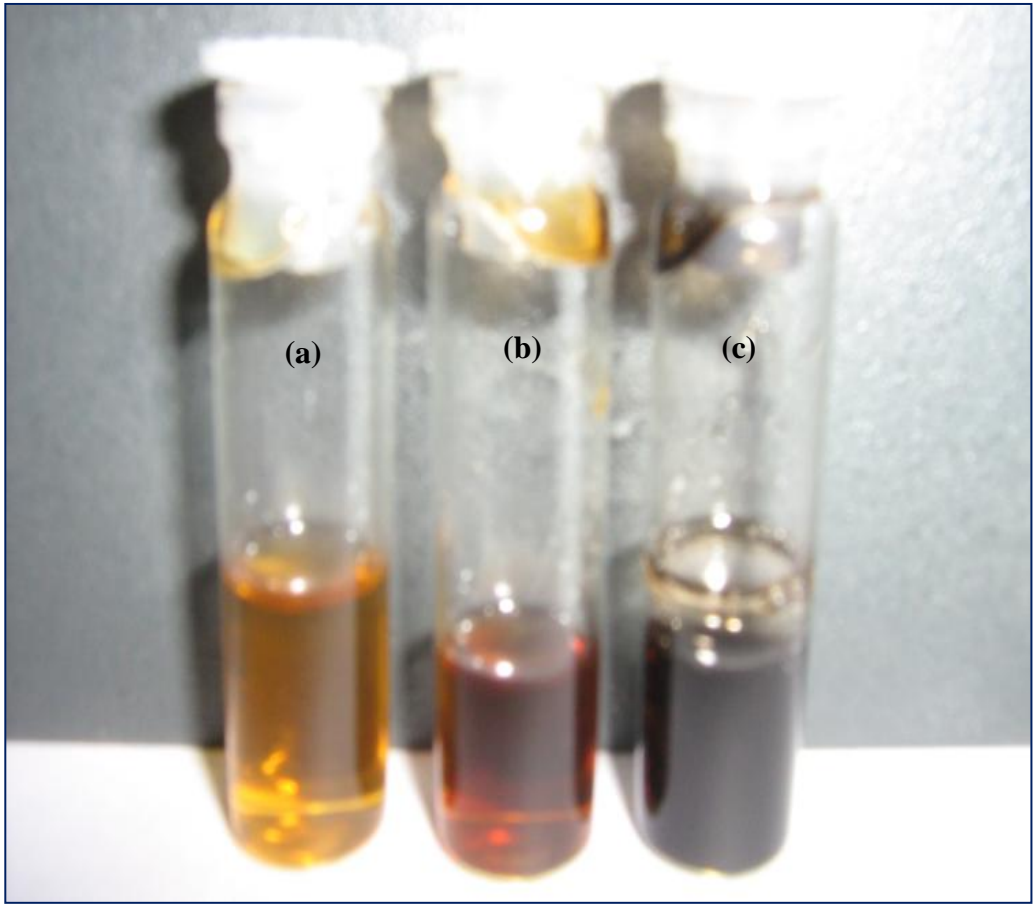

Fig 6

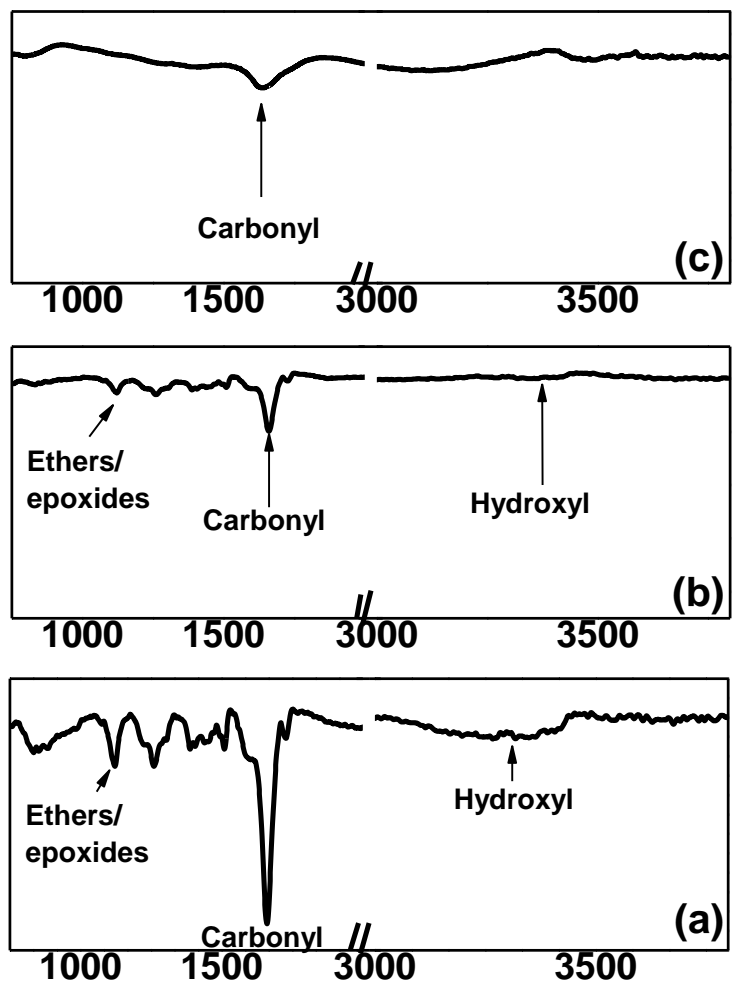

Fig 7

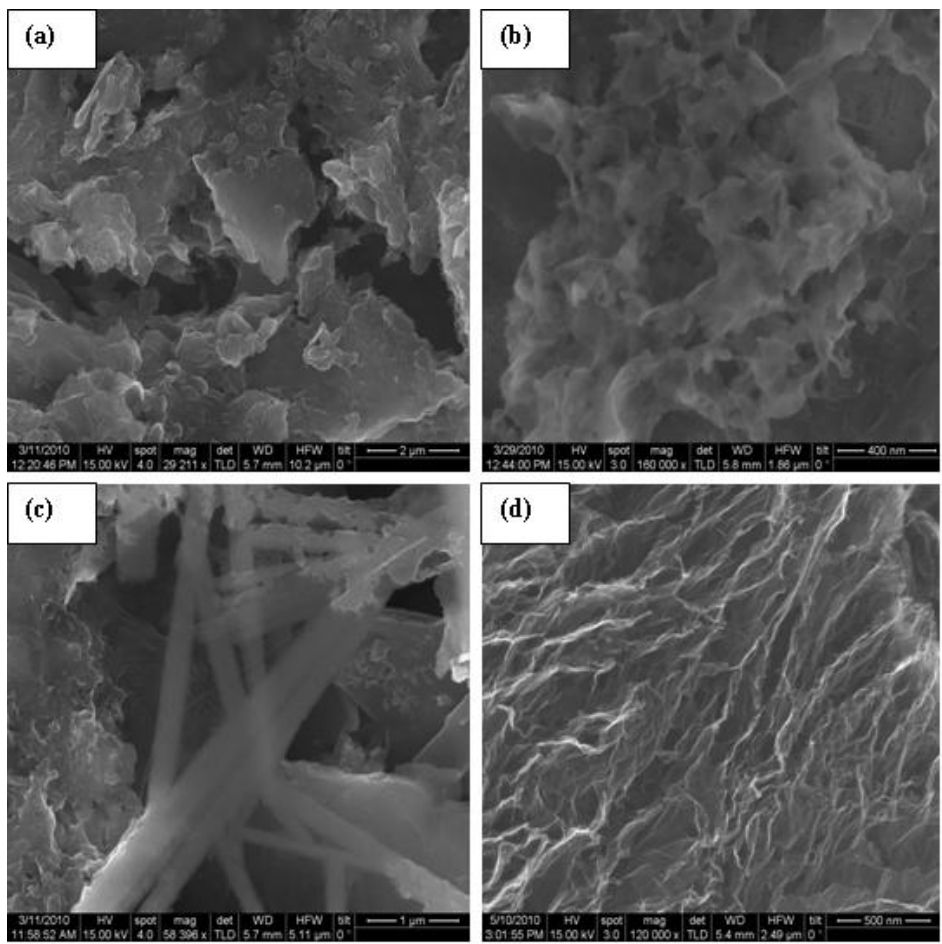

Fig 8

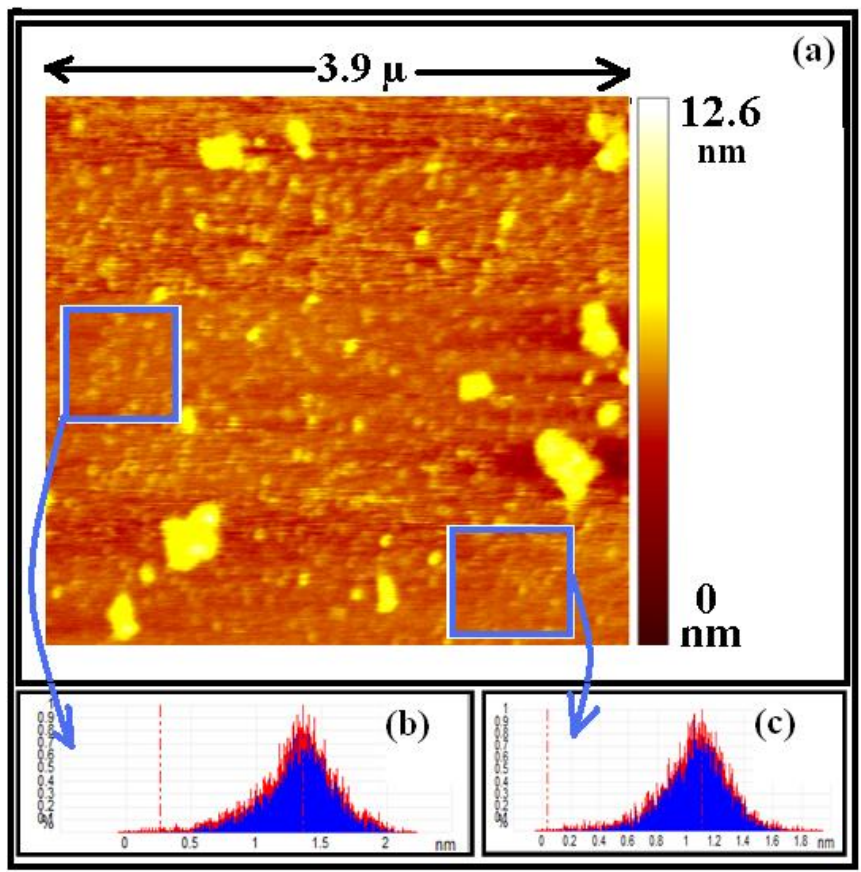

**Fig 9**

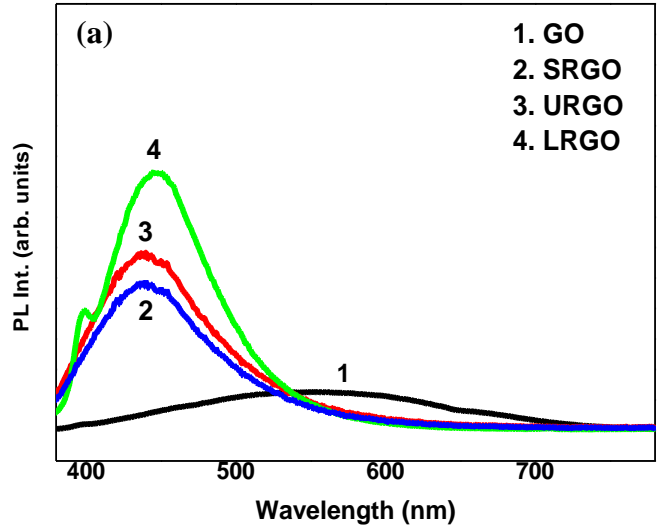
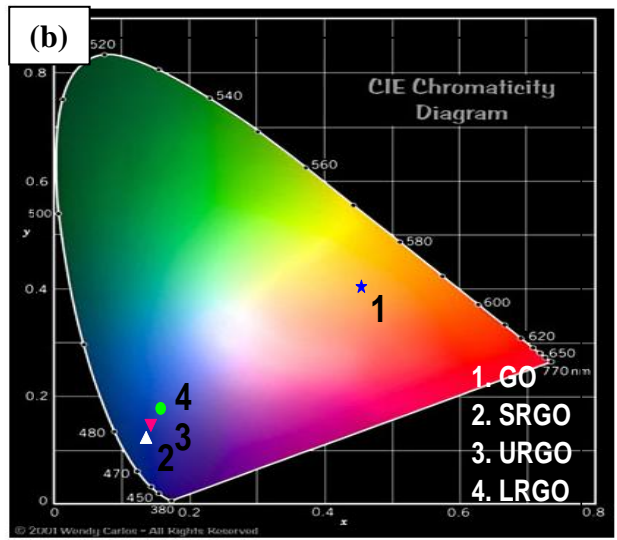

**Fig 10**

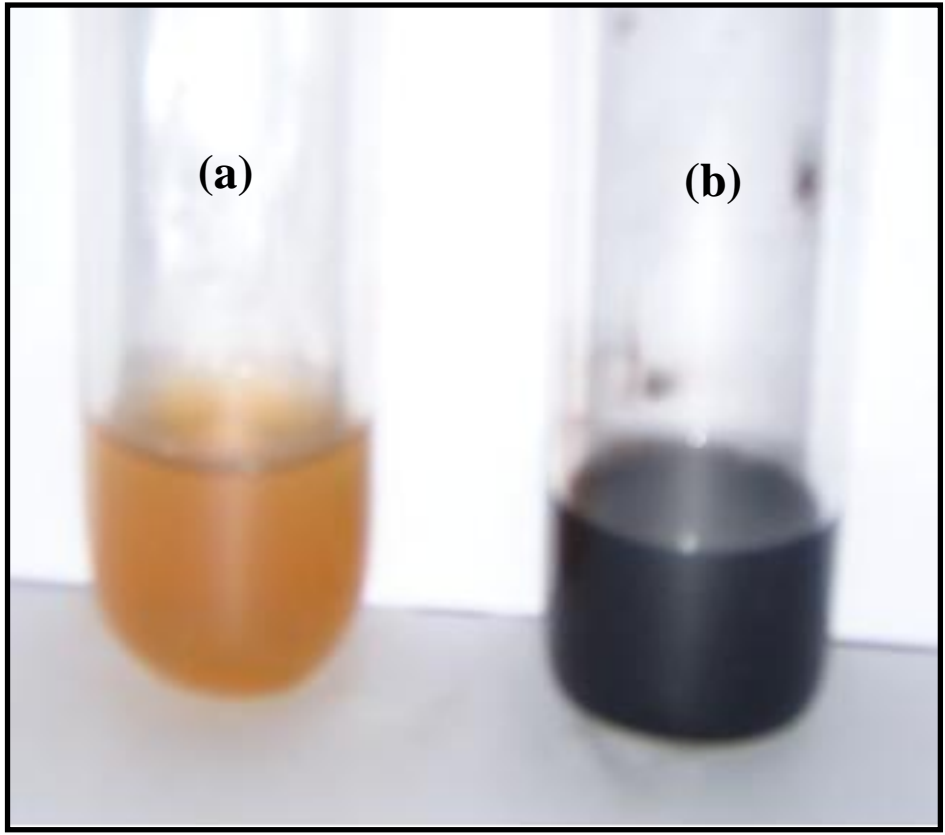

Fig 11

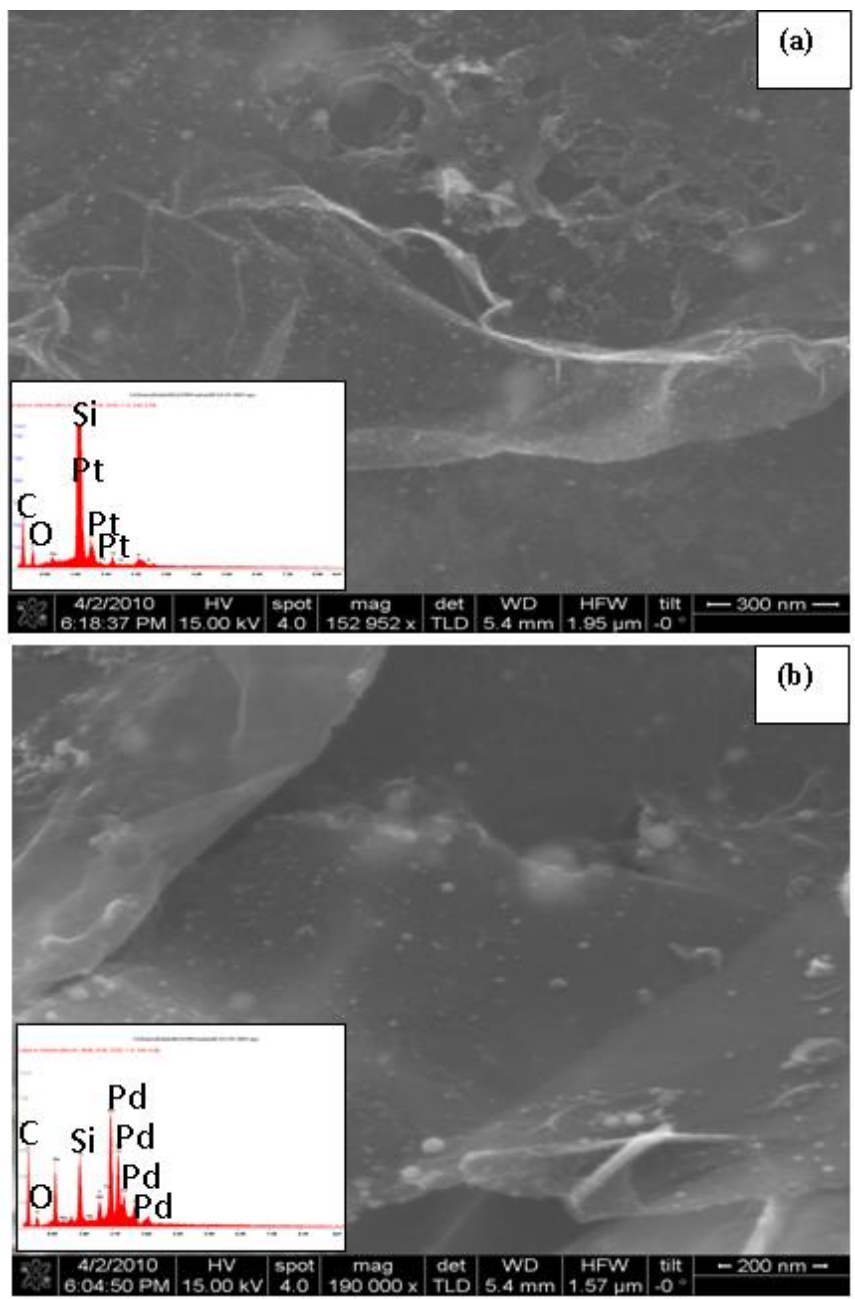

**Fig 12**